\documentclass[%
 reprint,
 superscriptaddress,
 onecolumn,
 amsmath,
 amssymb,
 10pt,
 aps,
 pra,
 citeautoscript,
 notitlepage,
 showkeys,
]{revtex4-1}

\usepackage{graphicx}
\usepackage{xcolor}
\usepackage{dcolumn}
\usepackage{array}
\usepackage{bm}
\usepackage[utf8]{inputenc}
\usepackage{algorithm}
\usepackage{algpseudocode}
\usepackage{fancyvrb}
\usepackage{newfloat}
\usepackage{booktabs}
\usepackage{dirtree}
\usepackage[%
  colorlinks=true,
  allcolors=blue
]{hyperref}
\usepackage{listings}
\DeclareFloatingEnvironment[placement={!ht},name=Listing]{listing}
\begin{document}

\title{
\texttt{maze}: Heterogeneous Ligand Unbinding along Transient Protein Tunnels
}

\author{Jakub Rydzewski}
\email[E-mail: ]{jr@fizyka.umk.pl}
\affiliation{Institute of Physics, Faculty of Physics, Astronomy and
Informatics, Nicolaus Copernicus University, Grudziadzka 5, 87--100 Torun,
Poland}


\begin{abstract}\noindent
  Recent developments in enhanced sampling methods showed that it is possible 
  to reconstruct ligand unbinding pathways with spatial and temporal resolution 
  inaccessible to experiments. Ideally, such techniques should provide an 
  atomistic definition of possibly many reaction pathways, because crude 
  estimates may lead either to overestimating energy barriers, or inability to 
  sample hidden energy barriers that are not captured by reaction pathway
  estimates. Here we provide an implementation of a new method
  [J. Rydzewski \& O. Valsson, J. Chem. Phys. {\bf 150}, 221101 (2019)] dedicated entirely to sampling
  the reaction pathways of the ligand-protein dissociation process. The program,
  called \texttt{maze}, is implemented as an official module for PLUMED 2, an 
  open source library for enhanced sampling in molecular systems, 
  and comprises algorithms to 
  find multiple heterogeneous reaction pathways of ligand unbinding from proteins 
  during atomistic simulations. The \texttt{maze} module requires only a 
  crystallographic structure to start a simulation, and does not depend on many 
  \textit{ad hoc} parameters. The program is based on enhanced sampling and 
  non-convex optimization methods. To present its applicability and flexibility, 
  we provide several examples of ligand unbinding pathways along transient 
  protein tunnels reconstructed by \texttt{maze} in a model ligand-protein system, 
  and discuss the details of the implementation.
  \\[0.1cm]
  \textbf{Manuscript version 1.0 release.} See 
  \url{https://arxiv.org/abs/1904.03929} for all versions.
  \\[0.1cm]
  \textbf{Date.} \today.
\end{abstract}


\maketitle


\section*{\label{sec:summary}Program Summary}

\begin{table}[ht!]
\begin{tabular}{p{4cm}p{1cm}p{10cm}}
  Program title:
    && \texttt{maze} \\

  Program version:
    && 1.0 \\

  Program files DOI:
    && \url{http://dx.doi.org/10.17632/x5zsgzxcnx.1} \\

  Licensing provisions:
    && L-GPL-3.0 \\

  Programming language:
    && \texttt{C++}11 \\

  Nature of problem:
    && Finding heterogeneous unbinding reaction pathways of ligand transport 
    processes such as dissociation and diffusion in ligand-protein systems 
    during molecular dynamics simulations. \\

  Solution method:
    && Enhanced sampling molecular dynamics method with non-convex optimization
    techniques performed on-the-fly during biased simulations. \\

  Unusual features:
    && \texttt{maze} is a module for the PLUMED 2 software, it must be used
    in conjunction with this software, embedded in a molecular dynamics code
    such as Gromacs, NAMD, LAMMPS, OpenMM, and Amber. \\

  Additional comments:
    && Program Github repository: \url{https://maze-code.github.io} \\

\end{tabular}
\end{table}


\section{\label{sec:intro}Introduction}

Atomistic simulations such as molecular dynamics (MD) provide sufficient temporal 
and spatial resolution to study complex physical processes. However, despite 
substantial progress in method development, conventional MD simulations are 
unable to access the longtime scales on which infrequent events 
occur~\cite{rydzewski2019finding,valsson2016enhancing,rydzewski2017ligand,rydzewski2017ligandb,
noe2018machine}. This so-called ergodic breakdown is associated with high energy 
barriers separating important energy minima. If these barriers are much higher 
than the thermal energy $k_\mathrm{B}T$, the system will be kinetically trapped 
in a metastable energy minimum during the course of simulations. To access longer 
time scales and reach the regime of high energy barriers of rare events enhanced 
sampling methods are needed. Finding collective variables (CVs) and estimating 
reaction pathways for rare events in a crude manner often leads to the 
overestimation of energy barriers, and thus, underestimation of exponentially 
dependent kinetic rates, which arise from inability to capture intrinsic degrees 
of freedom. This is related to the degeneracy of microscopic configurations 
originating from sampling wrong CVs and reaction pathways, which is likely to 
mask hidden energy barriers.

A typical example of an event that occurs on the long time scales is ligand-protein 
dissociation. The main computational hurdle, which makes reaction pathways for 
ligand-protein unbinding difficult to sample during the conventional MD
simulation, stems from accounting for transient internal features of proteins, 
such as tunnels. However, these features are important for mechanisms, 
thermodynamics, and kinetics of ligand unbinding, as the structural flexibility 
of tunnels and channels allows proteins to facilitate binding by adapting to 
ligands along possibly multiple pathways to the binding site. This transient 
dynamics poses a severe challenge even for enhanced sampling methods that have 
been used to sample reaction pathways in ligand unbinding so far. For instance, 
such methods either do not account for protein dynamics, interpolate reaction 
pathways linearly between bound and unbound states~\cite{heymann2000dynamic}, or 
sample protein tunnels by employing simple heuristics as random 
walks~\cite{ludemann2000substrates}.

New enhanced sampling methods, however, are often not used in the wider community 
because easy to use implementations are not available. A few of open source platforms 
serving as plugins to MD engines, e.g., PLUMED 2~\cite{tribello2014plumed,
bussi2018analyzing}, MIST~\cite{bethune2018mist}, SSAGES~\cite{sidky2018ssages}, 
and i-PI~\cite{kapil2018pi} are available. These open source platforms have 
significantly lowered the technical barrier for the application of advanced 
sampling methods in MD simulations. Considering the potential impact that enhanced 
sampling-based methods have on atomistic simulations, it is of considerable interest 
to develop new methods for already existing open source platforms that serve as the 
interface between enhanced sampling and atomistic simulations packages, e.g., 
LAMMPS~\cite{plimpton1995fast}, Gromacs~\cite{abraham2015gromacs}, or
NAMD~\cite{phillips2005scalable}. The contribution of this work is to provide 
the implementation of the \texttt{maze} code which can be interfaced with many MD 
engines via PLUMED 2. We provide a specific-use package dedicated entirely to 
simulating ligand-protein unbinding. Given the utilities provided by \texttt{maze}, 
running MD simulations of ligand unbinding is much easier through a simple input 
file for PLUMED 2, and enables enormous flexibility in creating new components 
for the method such as biasing potentials and optimizers.

The manuscript is organized as follows. In Sec.~\ref{sec:method}, the theoretical 
framework of the \texttt{maze} software is provided. We show in detail how 
ligand-protein interactions are calculated, and further optimized using a 
coarse-grained CV for the adaptive bias potential. Next, we provide a protocol 
showing how the bias potential is used in \texttt{maze}. We focus on the software 
and technical detail of the implementation in Sec.~\ref{sec:software}, including 
data preparation, numerical examples, and availability. Finally, we conclude our 
work in Sec.~\ref{sec:conclusions}.


\section{\label{sec:method}Method}

The method, which is the base component of \texttt{maze}, is able to find multiple 
diverse reaction pathways of ligand unbinding along transient protein 
tunnels~\cite{rydzewski2019finding}. In this section, we describe how this can be 
achieved. The method does not require an initial guess of intermediates nor a 
unbound state, which is very important, as many existing methods for calculating 
reaction pathways rely on it. Its only prerequisite is the knowledge of the X-ray 
binding site the ligand resides within. Since the searching procedure is performed 
iteratively during MD simulations, the method takes into account protein dynamics 
which is important to observe the openings and closings of transient tunnels and 
exits while probing the time scale on which these conformational changes occur.

The method relies on the following concepts:
\begin{itemize}

  \item[--] \textbf{Carefully selected CV describes interactions in a ligand-protein
    system}. In our protocol, this CV is optimized on-the-fly during the MD 
    simulation of ligand unbinding. Interestingly, as this part of the method 
    resembles many similarities with machine learning~\cite{goodfellow2016deep,mehta2019high}, 
    the CV can be seen as a specific loss function tailored to the optimization 
    problem of ligand unbinding, in the sense that a loss function is a function 
    that maps an event onto a real number intuitively representing some ``cost'' 
    associated with that event. Thus, we will use these two names interchangeably 
    throughout this article (Sec.~\ref{sec:lf}).

  \item[--] \textbf{Optimization method seeks to minimize this loss function.} 
    A non-convex optimization method is used to minimize this loss function 
    with specific constraints that are based on the surrounding protein tunnel or 
    channel. These constraints define a local CV space to sample ligand 
    conformations within the protein tunnel. In this manner \texttt{maze} learns
    the protein tunnels accessible for ligand transport (Sec.~\ref{sec:o}).

  \item[--] \textbf{Adaptive biasing potential enforces ligand dissociation.} 
    An adaptive bias potential is employed to enforce the transition 
    between the ligand current conformation and a localized minimum of the CV 
    in the following MD simulation steps. Once the bias reaches the computed
    minimum of the loss function, and if the ligand is still within the protein
    matrix, \texttt{maze} repeats the cycle (Sec.~\ref{sec:ab}).

\end{itemize}

In the following subsections we describe the aforementioned concepts in detail,
and show how each component of the method is connected to each other.


\subsection{\label{sec:lf}Loss Function for Ligand Unbinding}
Let consider a ligand-protein system ${\bf R}$ consisting of a $3n$-set of 
ligand coordinates ${\bf x}\equiv({\bf x}_1, \dots, {\bf x}_{3n})$ and a $3N$-set 
of protein coordinates ${\bf y}\equiv({\bf y}_1, \dots, {\bf y}_{3N})$. A CV that 
successfully models ligand unbinding needs to fulfill several important 
characteristics that are based on the notion that the problem of finding reaction 
pathways of ligand unbinding can be solved using optimization 
techniques~\cite{rydzewski2019finding}. The loss function needs to:

\begin{itemize}

  \item[--] \textbf{Tend to infinity as the ligand moves too close to the protein.} 
    This prohibits the method from sampling ligand configurations that would 
    clash with the protein.

  \item[--] \textbf{Decrease as the ligand unbinds from the protein.} 
    This provides a coarse estimate of how deeply ligand conformations are 
    buried within the protein tunnels.

\end{itemize}

Such loss functions can be used to obtain ligand unbinding pathways through
iterative minimization within MD simulations, however, we note that selecting a 
particular CV for this may depend on the decision if one wants to study 
effective interaction energy in a ligand-protein system, or use an approximate 
formula fulfilling the listed criteria. We note that in any case the reference 
ligand unbinding pathways should be used as an initial guess for methods like 
metadynamics~\cite{laio2002escaping,barducci2008well} or variationally enhanced 
sampling~\cite{valsson2014variational} if thermodynamic and kinetic properties 
of the system are to be reconstructed. 

In the current version \texttt{maze} implements an exponential loss function that 
can be used to describe ligand unbinding reaction pathways:
\begin{equation}
\label{eq:1}
  \mathcal{L}=\sum_{i=1}^{N_p}r_i^{-\alpha}\operatorname{e}^{-\beta r_i^{-\gamma}},
\end{equation}
where $r_i=\lambda|{\bf x}_k-{\bf y}_l|$ is the rescaled distance between atoms of 
the ligand and the protein of the $i$th pair, $N_p$ is the number of 
ligand-protein atom pairs, and $\theta=(\alpha, \beta, \gamma)$ are positive scaling 
parameters (or hyperparameters). Several combinations of these parameters have 
been tested in our previous studies~\cite{rydzewski2019finding,rydzewski2018kinetics}. 
We underscore that the number of ligand-protein atom pairs $N_p$ is based on
recalculating the neighbor list between the ligand and the protein, and it varies 
during the unbinding process as the ligand neighborhood changes. 

The loss function is calculated in the ligand neighborhood, and thus it needs a 
particular type of bounds for sampling solutions only from this neighborhood. 
Clearly, the procedure should not sample possible ligand conformations in the 
full accessible conformational space, because that may render spurious 
trajectories, e.g., the next ligand conformation may be behind some steric barrier, 
and overcoming it would result in an nonphysical dissociation process. To this 
aim, we automatically estimate the ligand neighborhood during the optimization procedure as a 
sphere centered on the ligand conformation from MD, of radius calculated as the 
minimal distance between ligand-protein atom pairs,

\begin{equation}
\label{eq:2}
  r_s=\min_i r_i.
\end{equation}

For a schematic figure showing the ligand neighborhood, see the right panel of
Fig.~\ref{fig:1}. Using such adaptive constraints limits the search for a global
minimum of the loss function to an easily found local minimum in the
proximity of the ligand conformation, and allows to sample non-linear reaction
pathways in any ligand-protein system during enhanced sampling simulations.
Thus, the local neighborhood of the ligand depends only on the current ligand
position and the minimal distance between ligand-protein atoms, which can be
roughly described as the width of the sampled protein
tunnel~\cite{rydzewski2019finding}.

The loss function $\mathcal{L}$ resembles many similarities to coordination
number, and can be seen as a coarse-grained CV which describes how deep the 
ligand is buried within the protein matrix, or as an effective interaction 
energy of a ligand-protein complex. For an example of how the loss function behaves 
during an MD simulation, see Fig.~\ref{fig:3}e. The loss function identifies
intermediate states along unbinding pathways, and although may fluctuate during
simulations, should decrease as the ligand unbinds from the protein.
We note also that this CV can be biased during simulations by any CV-based enhanced sampling
method, e.g., metadynamics~\cite{laio2002escaping}. Similar CVs were used by
enhanced sampling method in ligand binding
studies~\cite{trang2013rna,spitaleri2018fast}.


\subsection{\label{sec:o}Finding Minima of the Loss Function}

The minimization of the loss function can be performed using any robust non-convex 
optimization method. There are examples of many approaches of using such techniques 
in atomistic simulations~\cite{hansmann2002global,rydzewski2015memetic,
rydzewski2018conformational}. The \texttt{maze} code implements several such methods 
to optimize the CV chosen for ligand unbinding in the local space near the ligand. 
The full list of the implemented optimizers with their corresponding keywords 
in PLUMED 2 is given below:

\begin{itemize}

  \item[--] \texttt{MAZE\_SIMULATED\_ANNEALING}: simulated 
    annealing~\cite{kirkpatrick1983optimization}; optimizes the biasing direction
    based on the dynamically-adjusted Metropolis-Hastings
    method. For detailed description, see Ref.~\onlinecite{rydzewski2019finding}.
  
  \item[--] \texttt{MAZE\_MEMETIC\_SAMPLING}: memetic
    algorithm~\cite{rydzewski2015memetic,rydzewski2016machine,
    rydzewski2017thermodynamics,rydzewski2018kinetics} with learning heuristics such 
    as stochastic hill climbing and an adaptive Solis-Wets search; performs 
    exhaustive search using evolutionary algorithms.

\end{itemize}

In these methods, the ligand is encoded as its center-of-mass position. The
sampling procedure performed during the optimization phase draws a random
translation vector for the ligand conformation from the current step in the MD
simulation. If the translated ligand position is preferable in terms of the 
loss function, it is taken as the next conformation to which the ligand needs to be bias
toward, as this direction of biasing results in the decreases of the loss
function. This sampling procedure is repeated during the optimization phase to
unveil the position of the lowest loss function value. At 
each step of MD, the center-of-mass difference between the MD ligand
conformation and the optimal one serves as the direction of ligand unbinding.

Apart from the above optimizers, the \texttt{maze} implements also standard
techniques for sampling ligand-protein dissociation that are currently used by
the community, and can be used instead of the optimizers:

\begin{itemize}

  \item[--] \texttt{MAZE\_RANDOM\_WALK}: random walk; steers the ligand unbinding in a random
    direction.

  \item[--] \texttt{MAZE\_STEERED\_MD}: steered MD~\cite{heymann2000dynamic}; steers the
    ligand unbinding in a predefined direction.

  \item[--] \texttt{MAZE\_RANDOM\_ACCELERATION\_MD}: random acceleration
    MD~\cite{kokh2018estimation};
    changes the direction of the ligand unbinding randomly if the ligand does 
    not overcome a predefined threshold distance.

\end{itemize}

The above methods are not optimizers per se, but they can be used as samplers
for the biasing direction. In contrast to the optimizers, which provide an
optimal direction of biasing, the standard methods do not provide optimal
solutions for the adaptive bias, but return some direction of biasing (given by
the above listing).


\subsection{\label{sec:ab}Adaptive Biasing of Reaction Pathways}

Let $\{{\bf R}'_t\}=(\{{\bf x}'_t\}, {\bf y}_t)$ denote sampled ligand-protein
conformations. Once the optimal ligand conformation 
${\bf R}^*_t=\min_{{\bf R}'_t} \mathcal{L}({\bf R}'_t)$ is calculated, the 
ligand is biased in the direction of the optimal solution at time $t$ ${\bf x}^*_t$ 
along a selected transient protein tunnel. This stage is performed by biasing 
the positions of the ligand atoms with an adaptive harmonic potential defined 
as:
\begin{equation}
\label{eq:3}
  V({\bf x}_t)=
    \alpha
    \left(
      wt - ({\bf x} - {\bf x}^*_{t-\tau})
      \cdot\frac{{\bf x}^*_t - {\bf x}_{t-\tau}}{\|{\bf x}^*_t-{\bf x}_{t-\tau}\|}
    \right)^2,
\end{equation}
where $w$ is the biasing rate, 
$\tau$ is the interval at which the loss function is minimized, and $\alpha$ is
a force constant. The bias potential given by Eq.~\ref{eq:3} 
is a generalization of a simple harmonic biasing potential introduced in 
Ref.~\onlinecite{heymann2000dynamic}, modified to be able to bias curvilinear unbinding
pathways.

\begin{figure}[htb!]
  \includegraphics[width=0.6\textwidth]{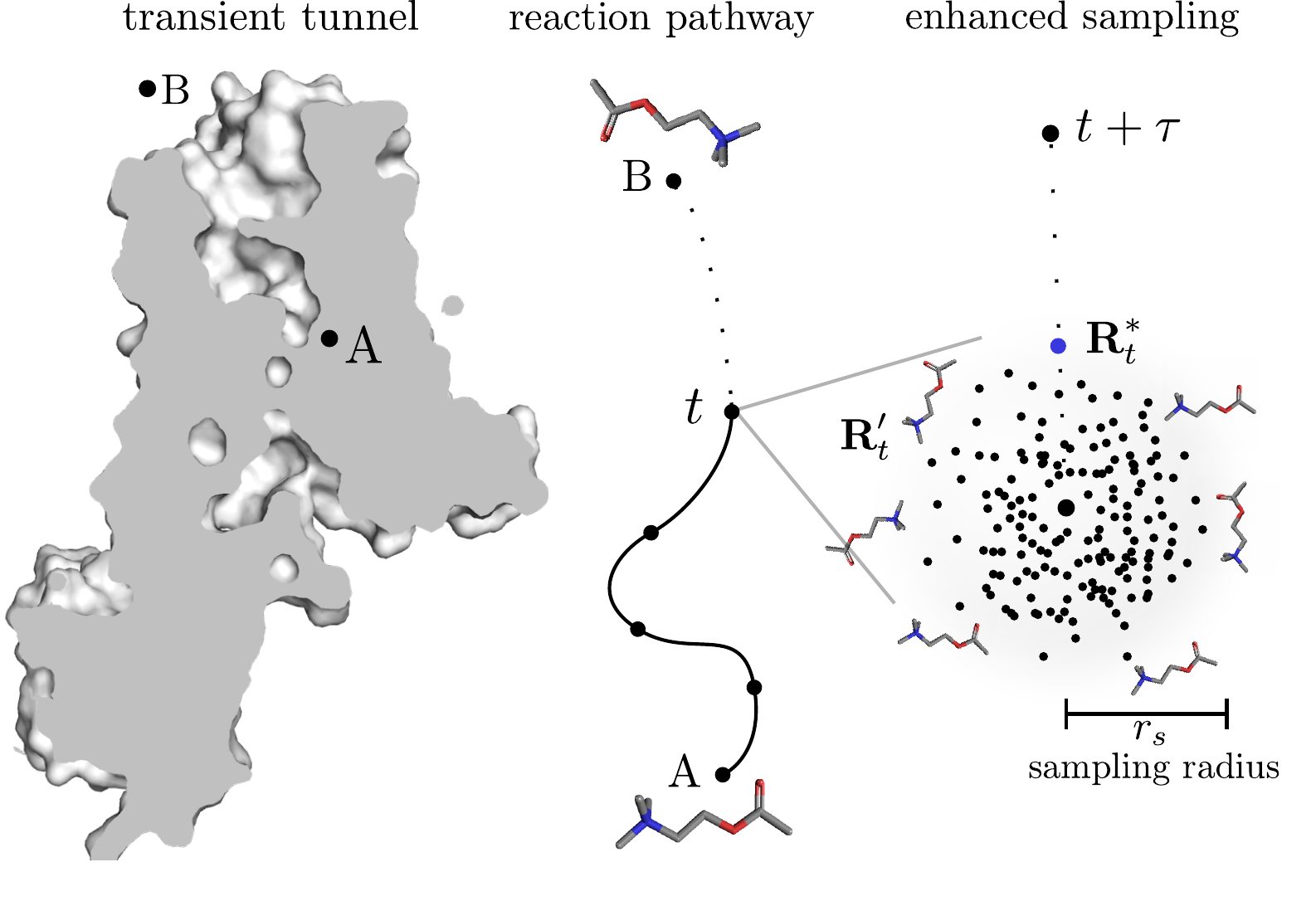}
\caption{Schematic depiction of the method implemented in the \texttt{maze}
  code. As an example, the unbinding of acetycholine from M1 muscarinic 
  receptor (PDB ID: 5cxv) is shown. The unbinding is initiated from the 
  bound state (A) of the M1-acetylcholine complex, and ends once the 
  ligand reaches solvent (B). The X-ray structure of M1 muscarinic 
  receptor indicates that there is a transient tunnel with a narrow gorge along 
  the exit route from A to B. The optimal direction of biasing is calculated by 
  minimizing the loss function between M1 muscarinic receptor and 
  acetylcholine. The right panel shows also how the sampling radius is
  calculated based on the distances between ligand and protein atoms.}
\label{fig:1}
\end{figure}

The biasing procedure is schematically depicted in Fig.~\ref{fig:1}, and
summarized as follows:
\begin{enumerate}

  \item Initialize the MD simulation and set time $t=0$.

  \item Sample ligand conformations $\{{\bf x}'_t\}$ within the protein 
    tunnel ${\bf y}_t$ using constraints defined by Eq.~\ref{eq:2}.

  \item Solve ${\bf R}^*_t=\min_{{\bf R}'_t} L({\bf R}'_t)$ using a
    non-convex optimization technique.

  \item Calculate a bias potential for the ligand ${\bf x}_t$ driving toward
    the optimal conformation ${\bf x}^{*}_t$ using Eq.~\ref{eq:3}.

  \item Run $\tau$ steps of the enhanced sampling simulation with the
    defined bias potential $V({\bf x}_t)$.

  \item Set $t=t+\tau$.

  \item Repeat the steps 2--5 during the MD simulation until $\mathcal{L}_t$ 
    reaches numerical zero when the ligand unbinds fully from the protein tunnel.

  \item Stop the MD simulation.

  \item Return putative reaction pathway coordinates $\{{\bf R}_t^*\}$ learned 
    during many optimization swaps.

\end{enumerate}
To find multiple unbinding pathways many simulations must be performed.

\section{\label{sec:software}Software}

\subsection{Installation}

The latest release of \texttt{maze} can be downloaded from the \texttt{maze} web
page~\cite{mazepage}. It is also possible to clone the
\texttt{maze} Github repository by appending the \texttt{maze} repository address 
to the command \texttt{git clone}. The \texttt{maze} code is an optional module 
of \texttt{maze} and thus it needs to be enabled when configuring the 
compilation with the \texttt{--enable-modules=maze} flag (or simply
\texttt{--enable-modules=all}) when running the configure script. Further 
information on compiling and installing PLUMED 2 can be found on in the PLUMED
documentation~\cite{plumeddoc}. The \texttt{maze} module is currently available in the
official development version of PLUMED 2, and will be released
in PLUMED 2.6 soon.


\subsection{Implementation}

\begin{figure}[htb!]
  \includegraphics[width=0.9\textwidth]{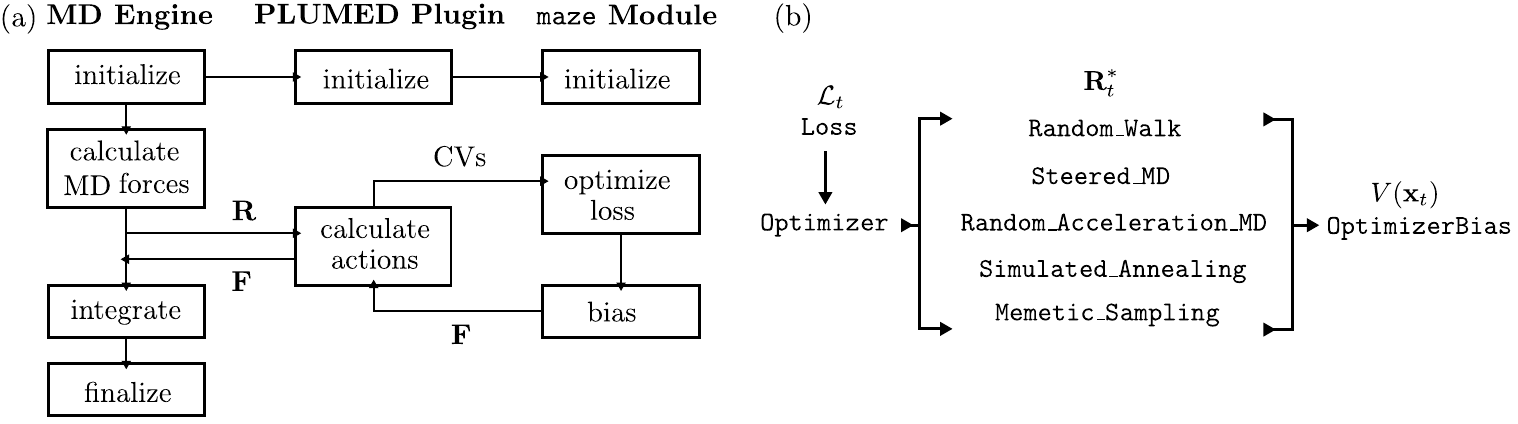}
  \caption{\texttt{maze} module for PLUMED. (a) Schematic flowchart of the 
  interfaces between an MD engine, the PLUMED
  plugin, and the \texttt{maze} module. A PLUMED input file defines directives
  for the module to be used. An MD engine shares information about the simulated
  system with the PLUMED plugin which calculates biasing forces ($\mathbf{F}$)
  based on the CVs dependent on the atomic coordinates ($\mathbf{R}$). The
  \texttt{maze} module optimizes the loss function, and biases a ligand toward
  the found minimum of the loss function. (b) Dependency scheme of the \texttt{maze} 
  module classes. See text for detailed descriptions.}
  \label{fig:2}
\end{figure}

Fig.~\ref{fig:2}a summarizes the workflow used in \texttt{maze} to perform a
ligand unbinding simulation. As \texttt{maze} is a module for PLUMED 2, it can 
be used with a wide range of MD codes, i.e., Gromacs and NAMD. Input files are
defined using the PLUMED input syntax. Within an input file every line is an
instruction for PLUMED to perform some action: calculating CVs, performing
analysis or running an enhanced sampling simulation. The PLUMED syntax as well
as all keywords are available in the PLUMED documentation~\cite{plumeddoc}.

The \texttt{maze} module comprises three base classes located within 
\texttt{src/maze} of the main directory of the installation (with the PLUMED 2 
base classes in parentheses):

\begin{itemize}

  \item[--] \texttt{Loss (Colvar)}: the loss function which act as the CV for 
    ligand unbinding,

  \item[--] \texttt{Optimizer (Colvar)}: the optimizer which provides a method 
    to minimize the loss function,

  \item[--] \texttt{OptimizerBias (Bias)}: the adaptive bias potential to enforce
    ligand-protein dissociation along transient tunnels.

\end{itemize}

The implementation is provided in the object-oriented paradigm of programming, and 
it is very easy to extend by adding new loss functions, optimizers (or standard
techniques), and biases. For a dependency scheme of the \texttt{maze} module, 
see Fig.~\ref{fig:2}b. Classes in \texttt{maze} can be extended by providing new
subclasses. The \texttt{Loss} and \texttt{Optimizer} classes are base classes
from which any derived class can inherit functions.
This setup allows for new ligand unbinding methods to be developed using a single 
unified interface, fully decoupled from MD codes. Every subclass
that derives from \texttt{Optimizer} must provide a \texttt{registerKeywords}
function that parses method-specific variables, and an \texttt{optimize}
function which is responsible for finding the optimal direction to bias ligand
unbinding. For instance, the \texttt{Random\_Walk}'s \texttt{optimize} function
returns a random unit vector and calculates the loss function to score a
particular ligand-protein configuration. The \texttt{optimize} functions of
\texttt{Steered\_MD} and \texttt{Simulated\_Annealing} return a predefined unit 
vector, and the optimal biasing direction calculated by simulated annealing,
respectively. In that sense, the \texttt{maze} module provides an unified
implementation of methods capable of calculating ligand unbinding reaction
pathways. For a schematic workflow of the \texttt{maze} module, see
Fig.~\ref{fig:2}a, which depicts how an MD engine is connected through the
PLUMED plugin to the \texttt{maze} module.

Regarding performance, the current \texttt{maze} version support inter-process
communication via OpenMP to parallelize the computation of the critical slower 
parts of the implementation such as calculating the loss function for each ligand 
and updating the neighbor list of ligand-protein atoms, which altogether render 
the sampling fast, as opposed to other implementations used currently for finding 
ligand exit pathways from proteins, for instance random acceleration MD implemented 
as a TCL script in NAMD.


\subsection{Data Preparation}

From the user perspective, before running the unbinding simulations using
\texttt{maze}, a chosen ligand-protein system of interest must be modeled.
Although PLUMED is interfaced with many MD codes, here we describe the
preparation of a ligand-protein complex only using Gromacs. For a very
useful tutorial for beginners regarding ligand-protein complexes, we redirect the
readers to Ref.~\onlinecite{lemkul2018proteins}. If the force-field parameters are not accessible
within a standard force field, we refer to a server for parametrizing small
organic molecules~\cite{dodda2017ligpargen}. Once the 
ligand-protein complex is modeled and ready for the production runs,
the user needs to define the loss function (List.~\ref{list:p1}) that is to be
biased during the simulations. Next, the optimizer must be defined
(List.~\ref{list:p2}), and then the biasing potential (List.~\ref{list:p3}). All
method-specific keywords and examples of usage are described and available in the
PLUMED documentation~\cite{mazedoc}.

The \texttt{maze} package requires very few \textit{ad hoc} parameters, with the 
main information that used needs to provide being the X-ray structure of a
ligand-protein complex. Ideally, the ligand conformation should be inferred from
the crystallographic structure, but if the crystallographic binding site with
the bound ligand is not know, then a docking procedure should suffice. Apart
from the structure, the user needs to provide a simple configurational
file which consists of the loss function (List.~\ref{list:p1}), the optimizer 
for the loss function (List.~\ref{list:p2}), and the adaptive bias potential for 
the optimizer (List.~\ref{list:p3}).

\begin{listing}
\caption{Example input file for a loss function labeled \texttt{loss} 
 with $\theta=(1,1,1)$.}
\label{list:p1}
\begin{lstlisting}
loss: MAZE_LOSS PARAMS=1,1,1
\end{lstlisting}
\end{listing}

In List.~\ref{list:p1} we show how to define a loss function. As shown in
Eq.~\ref{eq:1}, the loss function suitable for ligand unbinding simulations is
parametrized by the positive arguments, and they can be provided using the
\texttt{PARAMS} argument as a tuple. The defined loss function can be then
passed to the definition of an optimizer.

\begin{listing}
\caption{Some important settings in the parameter file for an optimizer 
  \texttt{opt} using simulated annealing as the minimization procedure, 
  which takes a loss function defined in List.~\ref{list:p1} as an 
  argument.}
\label{list:p2}
\begin{lstlisting}
MAZE_SIMULATED_ANNEALING ...
  LABEL=opt
  LOSS=loss
  N_ITER=1000
  OPTIMIZER_STRIDE=500000
  
  LIGAND=2635-2646
  PROTEIN=1-2634 

  PROBABILITY_DECREASER=300
  COOLING=0.95
  COOLING_SCHEME=geometric
... MAZE_SIMULATED_ANNEALING
\end{lstlisting}
\end{listing}

Aside from optimizer-specific parameters and the loss function, every 
optimizer should define the ligand and protein atoms using the \texttt{LIGAND} 
and \texttt{PROTEIN} keywords, the number of minimization steps (\texttt{N\_ITER}), 
and the stride (in MD steps) at which the optimization is launched during the 
simulation. Here, we used simulated annealing as an example optimizer to minimize the 
loss function. The initial value of the temperature-like parameter $T_0$ 
(\texttt{PROBABILITY\_DECREASER}) was modified according to the geometric cooling 
scheme: $T_j=kT_{j-1}$, where $j$ is the iteration number, and $k=0.95$
(\texttt{COOLING}). The defined optimizer should be then passed to the bias definition.
We used the input files shown in Lists.~\ref{list:p1}--\ref{list:p3} to run MD 
simulations of ligand unbinding.

\begin{listing}
\caption{Example input file for an adaptive bias potential which takes an
  optimizer \texttt{opt} defined in List.~\ref{list:p2}.}
\label{list:p3}
\begin{lstlisting}
MAZE_OPTIMIZER_BIAS ...
  LABEL=bias

  OPTIMIZER=opt
  ALPHA=3.6
  BIASING_RATE=0.02
... MAZE_OPTIMIZER_BIAS
\end{lstlisting}
\end{listing}

All questions regarding \texttt{maze} module for PLUMED can be asked using Gitter; 
for details see the Supporting Materials.


\section{Example Ligand Unbinding Pathways}

\subsection{Model Ligand-Protein System}
\begin{figure}[htb!]
  \includegraphics[width=0.9\textwidth]{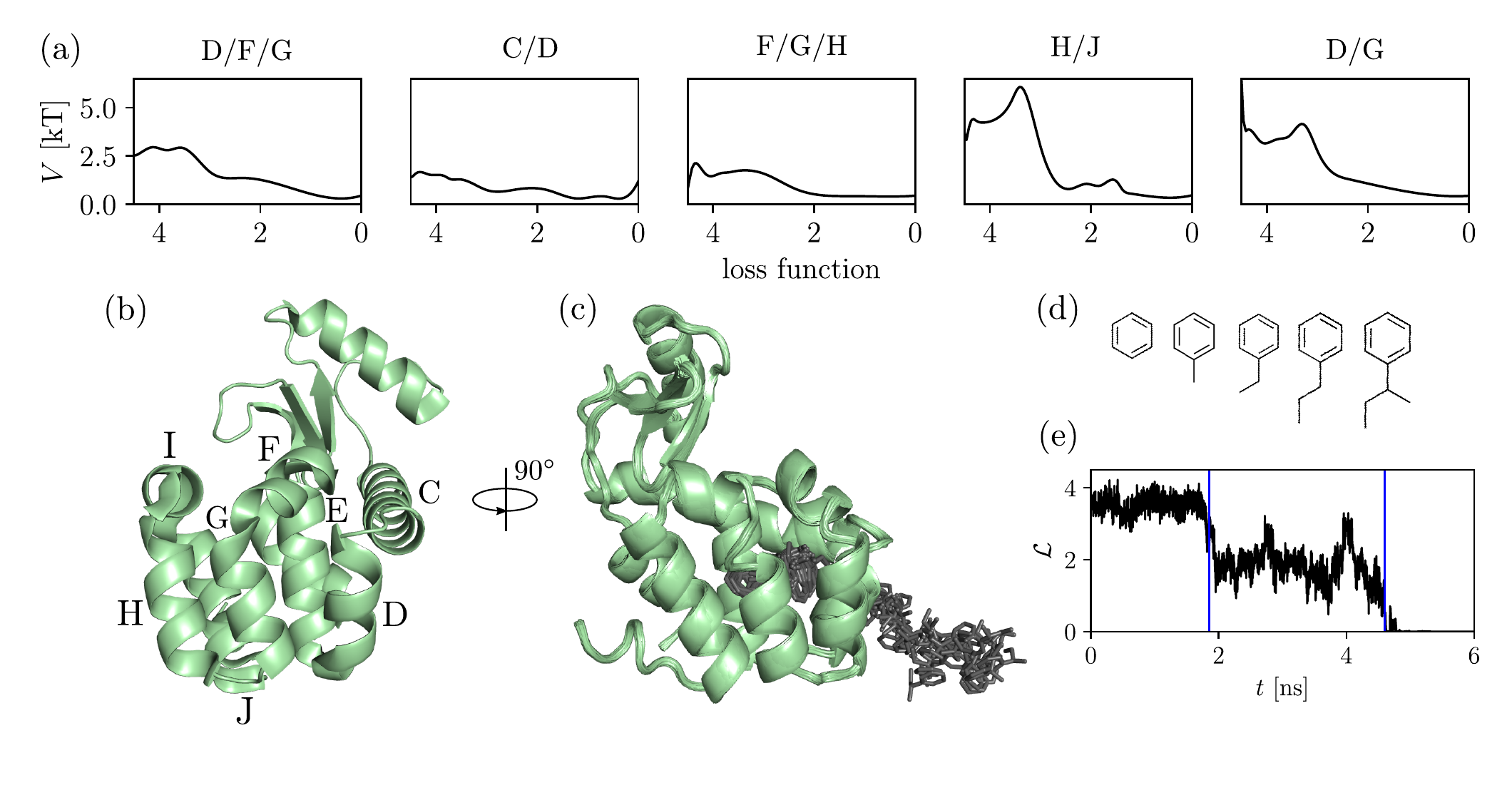}
  \caption{Structural and energetic characteristics of the ligand unbinding
  pathways of the T4L lysozyme L99A mutant. (a) Reproduced in part from 
  Ref.~\onlinecite{rydzewski2019finding}: Reaction pathways of the benzene unbinding from the 
  lysozyme L99A mutant classified in five clusters. The averaged bias potential
  used to force the transition from the bound to the dissociated state is shown
  as a nonlinear function of the minimized loss function (Eq.~\ref{eq:1}). (b) 
  The protein helices labeled as in Ref.~\onlinecite{rydzewski2019finding} that
  are used to name the ligand unbinding pathways. (c) An example of
  $sec$-butylbenzene unbinding pathway (D/G) from the protein matrix. (d) A
  series of congeneric homologous ligands (benzene, toluene, ethylbenzene,
  $n$-propylbenzene, $sec$-butylbenzene) that are dissociated from the T4
  lysozyme mutant sorted by size. (e) An example of the loss function
  parametrized by $\theta=(1,1,1)$, and calculated along an MD 
  trajectory. The loss function clearly identifies intermediates along the 
  unbinding pathway which is overlaid with corresponding ligand-protein
  conformations depicted in (c).}
  \label{fig:3}
\end{figure}

To show how the calculations done by the \texttt{maze} module are performed,
example ligand unbinding pathways from the T4 lysozyme L99A mutant (T4L) were
calculated for the following series of congeneric ligands bound to T4L: 
benzene, toluene, ethylbenzene, $n$-propylbenzene, and $sec$-butylbenzene
(Tab.~\ref{tab:2}). T4L is frequently used as a model system 
to study ligand unbinding from proteins (Fig.~\ref{fig:3}b). The ligands were
parametrized using the LigParGen server~\cite{dodda2017ligpargen} from the group of W. L.
Jorgensen. The OPLS/AA force field~\cite{kaminski2001evaluation} was used in the simulations; the SPC
water model~\cite{berendsen1981interaction} was used to solvate the system. The system was neutralized by
adding 6 CL ions. The modeling was performed using Gromacs
5.1.3~\cite{abraham2015gromacs}.

The MD simulations were launched using Gromacs 5.1.3~\cite{abraham2015gromacs} with a 2-fs time 
step. The system was simulated using periodic boundary conditions and the 
particle mesh Ewald method for long range
electrostatics~\cite{darden1993particle}. The system was first 
minimized, and then equilibrated in the NVT ensemble (10 ns) and in the 
NPT ensemble (10 ns) with the Parrinello-Rahman
barostat~\cite{martovnak2003predicting} set at 1 bar. 
The stochastic velocity rescaling thermostat~\cite{bussi2007canonical} was used to keep temperature 
at 300 K with T4L and benzene coupled. Bonds were constrained using
LINCS~\cite{hess1997lincs}. 
The unbinding simulations, in which the \texttt{maze} module was employed, 
were simulated in the NVT ensemble.

To efficiently calculate distances between the ligand and the
protein we used a parallelized neighbor list search which was recomputed every 
0.5 ps with a cut-off of 0.7 nm using OpenMP. The
optimization procedure was run every 200 ps using simulated annealing and
the loss function given by Eq.~\ref{eq:1}. The bias rate was set to $0.02 
~\text{\AA/ps}$ or $0.03~\text{\AA/ps}$ depending on the ligand size, and the biasing 
constant was $\alpha=3.6~\text{kcal/(mol~\AA)}$. To sample different ligand
unbinding pathways for the ligands (Fig.~\ref{fig:3}d), we run 200 10 ns
trajectories.

Our results show that the resulting ligand unbinding trajectories from the
T4L system can be classified into several different reaction pathways. Although
thermodynamic and kinetic quantities are important for studying ligand
unbinding, the structural representation of reaction pathways is as important. 
Clearly, the pathways show different structural mechanisms of the benzene unbinding, 
which is underlined additionally by the different values of the bias potential deposited along the
optimized loss function (Fig.~\ref{fig:3}a). Since, the detailed analysis of the benzene unbinding
pathways was published recently~\cite{rydzewski2019finding}, here we only focus
on a qualitative comparison of the reaction pathways. The sampled reaction pathways
are in very well agreement with recent computational 
studies. Namely, we found two unbinding pathways reported by Mondal et
al.~\cite{mondal2018atomic}, a dominant unbinding pathway found by Wang et al.
~\cite{wang2016mapping,wang2017biomolecular}, one pathway that resulted from
using Gaussian-accelerated MD~\cite{miao2015gaussian}, and four benzene escape
pathways found using temperature-accelerated MD~\cite{nunes2018escape}. For a
detailed analysis of benzene unbinding pathways from T4L, see
Ref.~\cite{rydzewski2019finding}.

Additionally, the ligand selectivity of choosing a particular T4L
pathway can be observed also for more complex then benzene ligands: toluene,
ethylbenzene, $n$-propylbenzene, and $sec$-butylbenzene. Our results
show that even for more complex ligands, the reaction pathways are very similar
to the benzene dissociation pathways (Tab.~\ref{tab:2}), which can be explained
intuitively as ligands similar in size to $sec$-butylbenzene are not enough to
change the state of the apolar T4L cavity from the closed to open
conformation~\cite{merski2015homologous}.

\begin{table}[h]
  \caption{Example unbinding pathways for a series of congeneric ligands bound to the lysozyme 
  L99A mutant. \textit{Note:} The results presented in this table are just for
  illustration purposes, and the correct number of the unbinding pathways from
  the protein may be higher as the number of the enhanced simulations run is
  relatively low comparing to Ref.~\onlinecite{rydzewski2019finding}. For
  questions about convergence in terms of the number of pathways,
  see Sec.~\ref{sec:discussion}.}
\label{tab:2}
\begin{tabular}{llcccl}
\hline
  PDB ID$^a$ & ligand & biasing rate [\AA/ps] & no. simulations & no. pathways & pathways \\
\hline
  4w52 & benzene  	        & 0.02 & 300 & 5$^b$ & D/F/G, C/D, F/G/H, H/J, D/G \\
  4w53 & toluene 		        & 0.02 & 50  & 4     & C/D, F/G/H, H/J, D/G        \\
  4w54 & ethylbenzene 		  & 0.03 & 50  & 4     & D/F/G, C/D, F/G/H, H/J      \\
  4w55 & $n$-propylbenzene  & 0.03 & 50  & 5     & D/F/G, C/D, F/G/H, H/J, D/G \\
  4w56 & $sec$-butylbenzene & 0.03 & 50  & 5     & D/F/G, C/D, F/G/H, H/J, D/G \\
\hline
  \multicolumn{4}{l}{$^a$ All the protein-ligand system are reported in
  Ref.~\onlinecite{merski2015homologous}.} \\
  \multicolumn{4}{l}{$^b$ Reported in Ref.~\onlinecite{rydzewski2019finding}.} \\
\end{tabular}
\end{table}


\section{\label{sec:discussion}Discussion}
\subsection{Related Methods}

While the residence time and binding free energy are a very important for 
ligand unbinding, an equally if not more important are the unbinding pathway adopted by
the ligand as the kinetic and thermodynamic quantities depend on the structural
definition of the ligand unbinding reaction pathways~\cite{pramanik2019can}.
Therefore, here we compare the method with other computational techniques 
that are able to sample ligand unbinding reaction pathways.

Tab.~\ref{tab:1} shows frequently
used CV-based enhanced sampling methods as well as newly introduced methods. 
The methods are compared by the MD simulation time
required to get a ligand unbinding reaction pathway and requirement for
obtaining such pathways. Methods like metadynamics (MTD), and its variant which
allows for calculating kinetic quantities for biased MD simulations, infrequent
MTD~\cite{wang2018frequency}, need an initial reaction pathway on which path-collective
variables~\cite{branduardi2007b} can be defined. Recently introduced
SGOOP~\cite{tiwary2016spectral} and
RAVE~\cite{ribeiro2018toward} require a rough estimate of several pre-selected CVs that define
a multidimensional initial pathway that is improved during simulations. A
different approach which combines biased MD simulations with long unbiased
trajectories is the transition-based reweighting analysis method
(TRAM)~\cite{wu2016multiensemble}.

Currently, a volume-based variant of MTD~\cite{capelli2019exhaustive} and the method 
implemented in the \texttt{maze}
module for PLUMED are capable of calculating ligand unbinding reaction pathways
with no requirement for an initial guess of the pathway. One important
distinction should be underlined here: the variant of MTD requires an extensive
post-processing procedure to recover unbinding pathways~\cite{capelli2019exhaustive}, unlike the method
presented here which gives a ligand unbinding pathway as a result of a biased MD
simulation. At this stage, however, \texttt{maze} is unable to calculate free
energy barriers along the pathways, which is not a problem for any MTD based
method, but only a rough estimate of energy barriers by calculating averaged biased
potential along the reaction pathways~\cite{rydzewski2019finding}. This averaged
bias potential, in the limit of slow pulling should give a correct energy barrier.
Therefore, we think that \texttt{maze} can serve as an ideal starting point for
MTD, SGOOP or RAVE.

\begin{table}[h]
  \caption{CV-based methods for computing ligand unbinding reaction pathways from 
  proteins. The MD simulation time depends on the properties of the ligand-protein 
  system studied as well as the set-up of the method used.}
\label{tab:1}
\begin{tabular}{lcllll}
\hline
  method & simulation time & requirements & reference \\
\hline
  MTD              & $\mu$s  & putative reaction coordinate for path-CVs       & Laio and Parrinello~\cite{laio2002escaping}       \\
  infrequent MTD   & $\mu$s  & putative reaction coordinate for path-CVs       & Wang et al.~\cite{wang2018frequency}              \\
  TRAM             & $\mu$s  & combining biased and long unbiased simulations  & Wu et al.~\cite{wu2016multiensemble}              \\
  SGOOP            &   ---   & initial guess of CVs monitored during test runs & Tiwary and Berne~\cite{tiwary2016spectral}        \\
  RAVE             &   ns    & putative reaction coordinate                    & Ribeiro and Tiwary~\cite{ribeiro2018toward}       \\
  volume-based MTD &   ns    &   ---                                           & Capelli et al.~\cite{capelli2019exhaustive}       \\
  {\tt maze}       &   ns    &   ---                                           & Rydzewski and Valsson~\cite{rydzewski2019finding} \\
\hline
\end{tabular}
\end{table}

\subsection{Parameters Transferability}
The parameters required by the optimization procedure in the \texttt{maze} module
are not sensitive, and therefore can be applied to other ligand-protein
complexes. This is because of the local conformational space of the diffused
ligand, which adapts as the ligand dissociates through the protein by estimating
the tunnel width. The local
space of the ligand is calculated automatically by Eq.~\ref{eq:2}, which makes
it adaptable to any ligand-protein system. The only parameters that need to be
selected before the production runs are the biasing rate $w$ introduced in
Eq.~\ref{eq:3} which defines how fast the biasing potential moves along the
optimized reaction pathway, and the interval at which the optimization is
re-launched during an MD simulation. Since the reaction pathways found
by the \texttt{maze} module are expressed as piecewise functions that are linear
over all intervals, the interval should be chosen as the minimal number of
unbinding direction re-calculations needed to approximate a curved reaction
pathway for a given ligand-protein complex.

\subsection{Convergence}
The convergence of the biased simulations can be tested ensuring the times of
ligand dissociation from a long-lived metastable bound state obey Poisson
statistics~\cite{salvalaglio2014assessing}.
Analysis of the unbinding time distribution obtained from the biased
simulations can be performed by comparing the Poisson cumulative distribution
function (CDF; $f(t)=1-\exp(t/\tau)$) with the empirical cumulative distribution 
function (ECDF) obtained from the unbinding probability distribution (i.e., the 
histogram of the number of unbinding events observed in the biased simulations over time)
from trajectories for each unbinding pathway~\cite{salvalaglio2014assessing}. 
The distance between CDF and ECDF 
which quantify their similarity can be calculated using a Kolmogorov-Smirnov test.
This procedure should be used to calculate the distance for every reaction
pathway found by the method, and, if the distances between ECDF and CDF are
small then no additional production runs are necessary. For an example of using
this analysis, we refer to Ref.~\onlinecite{rydzewski2019finding}.


\section{\label{sec:conclusions}Conclusion, Limitations, and Future Works}

We introduced the \texttt{maze} module for the PLUMED 2 software, which
implements a method to find multiple heterogeneous reaction pathways of ligand
unbinding from proteins. To the best of our knowledge, it is the first
contribution of enhanced sampling methods fully dedicated to ligand-protein
unbinding. The module is implemented in \texttt{C++} and interfaced to PLUMED 
2 as a new module, similarly to, e.g., the VES code~\cite{vescode}.
The method performs enhanced sampling of a coarse-grained variable that is
optimized during MD simulations to enforce ligand unbinding. The \texttt{maze}
code can be used with MD codes that can be interfaced with PLUMED 2, e.g.,
Gromacs, or NAMD.

In addition, we provided the detailed description of the sampling in
\texttt{maze} that requires using adaptive biasing and non-convex optimization.
We showed the performance of the \texttt{maze} module on a model ligand-protein
system, and analyzed reaction pathways for ligand unbinding from the T4
lysozyme L99A mutant, including input files needed to run the simulations.

As for the limitations of the method---we note that the bias potential can be 
used to estimate the energy barriers along the pathways only in a limited manner, 
and for a full thermodynamic description methods like 
metadynamics~\cite{laio2002escaping,barducci2008well}, or variationally 
enhanced sampling are needed~\cite{valsson2014variational}. At the current stage 
of development the \texttt{maze} package enables finding multiple heterogeneous 
ligand unbinding reaction pathways which can be used as an initial reaction pathway 
for more advanced enhanced sampling methods.

The \texttt{maze} module for PLUMED 2 provides a robust and efficient method to
enforce ligand unbinding, and by providing a good guess of ligand unbinding
pathways it may be used to limit the computational time spend to sample thermodynamic
and kinetic properties of complex physical systems  with other methods, also
provided by the PLUMED 2 software. In future work, we plan to focus on
using different bias potentials that would allow to compute free energy
estimates for each reaction pathway.

All the data and PLUMED input files required to reproduce the results reported
in this paper are available on PLUMED-NEST (\url{www.plumed-nest.org}), the
public repository of the PLUMED consortium~\cite{plumed}, as
\texttt{plumID:19.056}.


\begin{acknowledgments}
During its development \texttt{maze} was partially supported by the National Science
Center grants (2015/19/N/ST3/02171, 2016/20/T/ST3/00488, and 2016/23/B/ST4/01770). 
The example MD simulations were computed at the Interdisciplinary Center for 
Modern Technologies, Nicolaus Copernicus University, Poland. We would like to 
thank all the people who provided valued suggestions, and made time to discuss 
the code privately with the author. In particular, we thank O. Valsson, M.
Zieliński, and P. Różański for critically reading the article, G. Bussi for help
in including the \texttt{maze} module in the official PLUMED 2 release, and 
H. Grubm\"{u}ller, W. Nowak, and M. Parrinello for guidance.
\end{acknowledgments}


\bibliography{maze}

\end{document}